%% file: samplepaper.tex
\RequirePackage{amsmath}
\documentclass[runningheads]{llncs}
\usepackage[T1]{fontenc}
\usepackage{graphicx}
\usepackage{algorithm}
\usepackage{algpseudocode}
\usepackage{amssymb}
\usepackage{bm}
\usepackage{diagbox}
\usepackage{float}

\usepackage{booktabs}  
\usepackage{multirow}  
\usepackage{array}     
\usepackage{makecell}  
\usepackage{siunitx}   
\usepackage{hyperref}
\usepackage[capitalise]{cleveref}
\usepackage{subcaption}
\usepackage{marvosym}
\usepackage{color}

\begin{document}
\title{Sampling-Pattern-Agnostic MRI Reconstruction 
through Adaptive Consistency Enforcement
with Diffusion Model}
\titlerunning{SPA-MRI}
%
\author{Anurag Malyala\inst{1,2} \and
Zhenlin Zhang\inst{1,2} \and
Chengyan Wang\inst{3}\and
Chen Qin\inst{1}\textsuperscript{(\Letter)}}


%
\institute{Department of Electrical and Electronic Engineering \& I-X, Imperial College London, London, UK \\ \and 
Department of Computing, Imperial College London, London, UK\\ \and 
Human Phenome Institute, Fudan University, Shanghai, China  \\
\email{\{a.malyala23, zhenlin.zhang23, c.qin15\}@imperial.ac.uk \\ wangcy@fudan.edu.cn}\\
}
%
\maketitle              
\begin{abstract}

Magnetic Resonance Imaging (MRI) is a powerful, non-invasive diagnostic tool; however, its clinical applicability is constrained by prolonged acquisition times. Whilst present deep learning-based approaches have demonstrated potential in expediting MRI processes, these methods usually rely on known sampling patterns and exhibit limited generalisability to novel patterns. In the paper, we propose a sampling-pattern-agnostic MRI reconstruction method via a diffusion model through adaptive consistency enforcement. Our approach effectively reconstructs high-fidelity images with varied under-sampled acquisitions, generalising across contrasts and acceleration factors regardless of sampling trajectories. We train and validate across all contrasts in the MICCAI 2024 Cardiac MRI Reconstruction Challenge (CMRxRecon) dataset for the ``Random sampling CMR reconstruction'' task. Evaluation results indicate that our proposed method significantly outperforms baseline methods.
\end{abstract}
%
%
\section{Introduction}

Magnetic Resonance Imaging (MRI) is crucial in clinical diagnosis, offering high-resolution, radiation-free visualisation of internal structures. However, prolonged scanning times remain a significant limitation and can cause patient discomfort resulting in motion artefacts. While under-sampling data expedites scans, it breaks the Nyquist sampling limit, resulting in aliasing artefacts.

Recent advancements in machine learning (ML) have enabled accurate reconstructions from under-sampled MRI acquisitions, addressing the challenge of lengthy scan times. The reconstruction of MR images is usually framed as an inverse problem traditionally solved with various techniques, including compressed sensing\cite{Lustig}, parallel imaging, low-rank methods, etc. However, contemporary research has demonstrated that deep learning (DL)-based methods significantly outperform these conventional techniques \cite{mardani2018deep}, thus garnering increased attention in the field.

Deep convolutional neural networks (CNNs) are the predominant DL models for under-sampled MRI reconstruction. Schlemper et al. \cite{Schlemper} combined data consistency with CNNs to enhance reliability and accuracy in image reconstruction. Qin et al. \cite{Qin} further refined this by implementing a Convolutional Recurrent Neural Network (CRNN) to exploit temporal redundancies in dynamic MRI sequences. Guang et al. \cite{8233175} suggested improving the image fidelity by applying a conditional Generative Adversarial Network (GAN) \cite{goodfellow2014generativeadversarialnetworks} as a further refinement step. Aggarwal et al. \cite{aggarwal2018modl} proposed a model-based DL approach (MoDL) for parallel imaging data, decomposing it into a denoising and data consistency step optimised by conjugate gradient (CG) method.

The recent advancement in generative models has introduced diffusion models \cite{Ho,Song,Song2} to MRI reconstruction. Wherein two main approaches have emerged: 
Conditional methods as proposed by \cite{Xiang} and \cite{gungor2023adaptive}, which apply a conditional Denoising Diffusion Probabilistic Model (DDPM)\cite{Ho} framework, and unconditional methods, which utilise posterior guidance for reconstruction-based inverse problems. The latter, exemplified by Chung et al. \cite{chungdps}, Peng et al. \cite{Peng}, and Song et al. \cite{song2022solvinginverseproblemsmedical}, are not constrained to fixed measurement models, leading to their increasing prominence in reconstruction tasks. In their recent work, Chung et al. \cite{chung2023dds} extended this to parallel imaging following the lines of \cite{aggarwal2018modl} with Decomposed Diffusion Sampling (DDS).

Conditional methods perform well in known settings but fail to generalise to new sampling patterns, while unconditional models offer flexibility at the cost of lengthy inference and tuning. In this work, we propose ``Sampling Pattern Agnostic MRI Reconstruction (SPA-MRI)'', a diffusion-based framework that robustly reconstructs high-quality scans across diverse contrasts, generalising over sampling trajectories and acceleration factors. SPA-MRI achieves faster inference with adaptive hyperparameters, significantly outperforming baseline methods in evaluations.
\section{Preliminaries}

\subsection{Diffusion Models}
Diffusion models \cite{Ho,Song,Song2} form a class of generative models which learn to transform a known distribution into a target empirical distribution in a step-by-step process of noising and denoising. The noising process can be represented as an It\^o stochastic differential equation (SDE) as described by Song et al. \cite{Song}:
\begin{equation}
    d\textbf{x} = \textbf{f}(\textbf{x},t)dt + g(t)d\textbf{w},
\label{SDE}    
\end{equation}
where $\textbf{x}$ is the data degraded over time $t \in [0, T]$. $\textbf{f}( \cdot , t)$, $g(t)$ are the drift and diffusion coefficients, and $\mathbf{w}$ is the Wiener process. $\textbf{f}( \cdot , t)$ and $g(t)$ are defined such that at $t=0$,  $x_0 \sim p_{\text{data}}$ which is the data distribution and at $t=T$, $ x_T \sim p_T = \mathcal{N}(0,\mathbf{I})$, a tractable distribution. One parametrisation of this is the Variance Preserving (VP) SDE:
\begin{equation}
    \label{vpsde}
    d\mathbf{x} = -\frac{1}{2}\beta(t)\mathbf{x}dt + \sqrt{\beta(t)}d\mathbf{w},
\end{equation}
where $\beta(t)$ is a monotonically increasing function that controls the noise scale. Song et al.\cite{Song} have demonstrated that DDPM\cite{Ho} are equivalent to the setting of VP-SDE, which is used in this research.

To sample from the data distribution $p_0$, we solve the corresponding reverse VP-SDE \cite{anderson1982reverse} for \cref{vpsde}:
\begin{equation}
    d\textbf{x} = [-\frac{1}{2}\beta(t)\textbf{x} - \beta(t)\nabla_{\textbf{x}_t}\log p_t(x_t) ] dt + \sqrt{\beta(t)}d\Bar{\textbf{w}}.
\label{reSDE2} 
\end{equation}
In the above reverse SDE,  $\nabla_{x_t} \log (p_t(x_t))$ represents the score function of the distribution, which is estimated with a neural network $s_\theta(\mathbf{x}, t)$, and $\Bar{\textbf{w}}$ is the Wiener process in reverse.



\subsubsection{Denoising Diffusion Implicit Models}
\label{subsec:ddim}
Given the iterative nature of SDEs and DDPMs, sampling is inadvertently slow. Song et al. \cite{song2020ddim} introduced a deterministic sampling strategy called Denoising Diffusion Implicit Models (DDIM), where they update the sampling rule as:
\begin{equation}
    \mathbf{x}_{t-1} = \sqrt{\Bar{\alpha}_{t-1}} \left( \mathbf{x}_0(\mathbf{x}_t) \epsilon_\theta(\mathbf{x}_t, t) \right) + \sqrt{1-\Bar{\alpha}_{t-1} - \eta^2\sigma^2}\epsilon_\theta(\mathbf{x}_t, t) + \eta\sigma\epsilon,
    \label{ddim_update}
\end{equation}
and
\begin{equation}
    \mathbf{x}_0(\mathbf{x}_t) = \frac{\mathbf{x}_t - \sqrt{1-\Bar{\alpha}_t}\epsilon_\theta(\mathbf{x}_t, t)}{\sqrt{\Bar{\alpha}_t}},
\end{equation}
where $\alpha_t = 1 - \beta_t$, $\Bar{\alpha}_t = \prod_{s=0}^t\alpha_s$, and $\sigma = \beta_t (1-\Bar{\alpha}_{t-1})/(1-\Bar{\alpha}_t)$. The parameter $\eta$ controls the stochasticity of the update, and $\epsilon_\theta(\mathbf{x}, t)$ is the denoising model for \cite{Ho}, and for DDIM, $\eta=0$. $\mathbf{x}_{t-1}$ is the prediction of the next denoised state from $\mathbf{x}_t$ and $\mathbf{x}_0(\mathbf{x}_t)$ is the current estimate for $\mathbf{x}_0$. This deterministic sampling strategy thus allows for faster sampling from the learned prior model. \footnote{Given the formulation under \cref{reSDE2}, the score model $s_\theta(\mathbf{x}_t,t)$ is equivalent to the noise model $\epsilon_\theta(\mathbf{x}_t, t)$ as $s_\theta(\mathbf{x}_t, t) \approx -\epsilon_\theta(\mathbf{x}_t, t)/\sqrt{1-\Bar{\alpha}_t}$}

\subsection{Diffusion Models for Medical Inverse Problems}
\label{subsec:diffusion_inverse}
Reconstructing MR images from under-sampled k-space data can be stated as a linear inverse problem:
\begin{equation}
y= \mathcal{A} x + \epsilon, 
\label{y_x_relation in MRI}
\end{equation}
where $\mathcal{A} = \mathcal{F}\mathcal{M}s^c$ comprises three key components: the Fourier encoding matrix $\mathcal{F}$, the under-sampling mask $\mathcal{M}$, and the coil-wise sensitivities $s^c$. $\mathbf{y}$ represents the observed under-sampled k-space measurement, while $\mathbf{x}$ is the fully sampled image we aim to reconstruct and $\epsilon$ is the measurement noise in k-space. 

To solve the inverse problem, we sample from the posterior distribution $p(\mathbf{x}|\mathbf{y})$. Using Bayes' theorem, we have $p(\mathbf{x}|\mathbf{y})=p(\mathbf{y}|\mathbf{x})p(\mathbf{x})/p(\mathbf{y})$, where the diffusion model serves as the data prior $p(\mathbf{x})$. This approach leads to an updated conditional for the reverse VP-SDE:
\begin{equation}
    d\mathbf{x} = [-\frac{1}{2}\beta(t)\textbf{x} - \beta(t)(\nabla_{\mathbf{x}_t}\log p_t(\mathbf{x}_t) + \nabla_{\mathbf{x}_t} \log p(\mathbf{y}|\mathbf{x}_t))] dt + \sqrt{\beta(t)}d\mathbf{w}.
\label{reSDE3} 
\end{equation}

It is noteworthy that the conditional score function $\nabla_{\mathbf{x}_t} \log p(\mathbf{y}|\mathbf{x}_t)$ is generally intractable. However, a few approaches to make this tractable have been proposed.  
For example, Chung et al. \cite{chungdps} proposed an efficient approximation, expressing the unknown as $p(\mathbf{y}|\mathbf{x}_t) \approx p(\mathbf{y}|\mathbf{x}_0) = \mathbb{E}[\mathbf{x}_0(\mathbf{x}_t)]$, resulting in following approximation of the desired score called Diffusion Posterior Sampling (DPS): 
\begin{equation}
    \label{dps-eq}
    \nabla_{x_t} \log p_t(x_t) + \nabla_{x_t} \log p_t(y|x_t) \simeq s_\theta(\mathbf{x},t) - \frac{1}{\sigma^2}\nabla_{\mathbf{x}_t}||\mathbf{y}-\mathcal{A}(\mathbf{x}_0(\mathbf{x}_t))||_2^2.
\end{equation}
This approach substitutes the unknown fully-sampled image with its current approximation, which is conditional only on $\mathbf{x}_t$. Subsequent works have confirmed that this approximation enables sampling from the true posterior distribution and can retrieve the true sample \cite{routpsld}.

In contrast to DPS, which requires model gradient calculations, Wang et al. \cite{wang2022zero} proposed the Denoising Diffusion Null-Space Model (DDNM). This method first performs unconditional sampling from $p(\mathbf{x}_t)$, then enforces consistency by back-projecting null-space errors:
\begin{equation}
    \label{ddnm-update}
    \mathbf{x}_0(\mathbf{x}_t)' = (\mathbf{I} - \mathcal{A}^T\mathcal{A})\mathbf{x}_0(\mathbf{x}_t) + \mathcal{A}^T\mathbf{y}.
\end{equation}

\noindent \textbf{Shortcut sampling for fast reconstruction}
Current methods like DPS \cite{chungdps} and DDNM \cite{wang2022zero} generate priors in the target distribution and enforce posterior constraints. However, they overlook seed value initialisation, defaulting to Gaussian noise as default initialisation. Recognising under-sampled data as degraded versions of fully sampled data, we can improve the initialisation. 
Prior works addressed this by applying the diffusion process forwards using DDIM na\"{i}vely \cite{song2020ddim}:
\begin{equation}
        \label{ddim-forward}
        \mathbf{x}_{t+1} = \sqrt{\alpha_{t+1}} \left( \mathbf{x}_0(\mathbf{x}_t)\epsilon_\theta(\mathbf{x}_t, t) \right) + \sqrt{1-\alpha_{t+1}}\epsilon_\theta(\mathbf{x}_t, t).
\end{equation}

While this preserves source information, the resulting samples can lack realism.



\section{Methodology}
Our proposed method \textbf{SPA-MRI} addresses key challenges in MRI reconstruction from under-sampled data. We mainly make three contributions: (1) an enhanced initialisation process, (2) improved data consistency for parallel imaging, and (3) a novel frequency decomposed posterior loss. These contributions collectively enforce consistency and improve reconstruction quality and efficiency while generalising across sampling patterns. Below, we detail each component of our methodology.

\input{methodology}

\section{Experiments}

\noindent\textbf{Data Preprocessing}
In our study, we train and evaluate on the CMRxRecon2024 challenge training dataset \cite{wang2024cmrxrecon2024}. The dataset comprises four distinct modalities: Cardiac cine, Aorta, Mapping, and Tagging. Since the ground truth for the test set are not public yet, we implemented a stratified split, allocating 70\% (981 files) for training and 30\% (423 files) for testing.

Our pre-processing pipeline addresses heterogeneous image dimensions by rescaling all training data to $256\times512$ using PyTorch's \cite{paszke2019pytorch} ``nearest-exact'' interpolation. While rescaling introduces some loss, we found this method had a negligible impact across all contrasts. Images are reverted to their respective original dimensions for the projection steps. Our model processes three consecutive frames simultaneously, using a 6-channel pseudo-real representation.  
Other preprocessing includes standardising to unit standard deviation and scaling to the maximum input magnitude, constraining values to the range $[-1, 1]$.

\vspace{+3pt}
\noindent\textbf{Implementation Details}
We implement our model based on the Improved Denoising Diffusion Probabilistic Model (iDDPM) \cite{nichol2021iddpm}. We train the model jointly on all modalities, using a learning rate of $10^{-4}$. The noise schedule $\beta(t)$ has 4000 discretisation steps with a cosine $\beta$ schedule. For inference, we use 200 DDIM steps and 25 inversion steps, with a back projection base scale $\xi$ of 3. The centre $32\times32$ k-space region is used in the low frequency term in \cref{freq_decomp} with weights $\lambda_1 : \lambda_2 = 0.4:0.6$. A comprehensive table of all parameters is provided in the supplementary material.\footnote{Reproducibilty Note: The inference code and model weights will be made public after the conclusion of the CMRxRecon2024 challenge. }

\begin{figure}[!thbp]
    \centering
    \includegraphics[width=\linewidth]{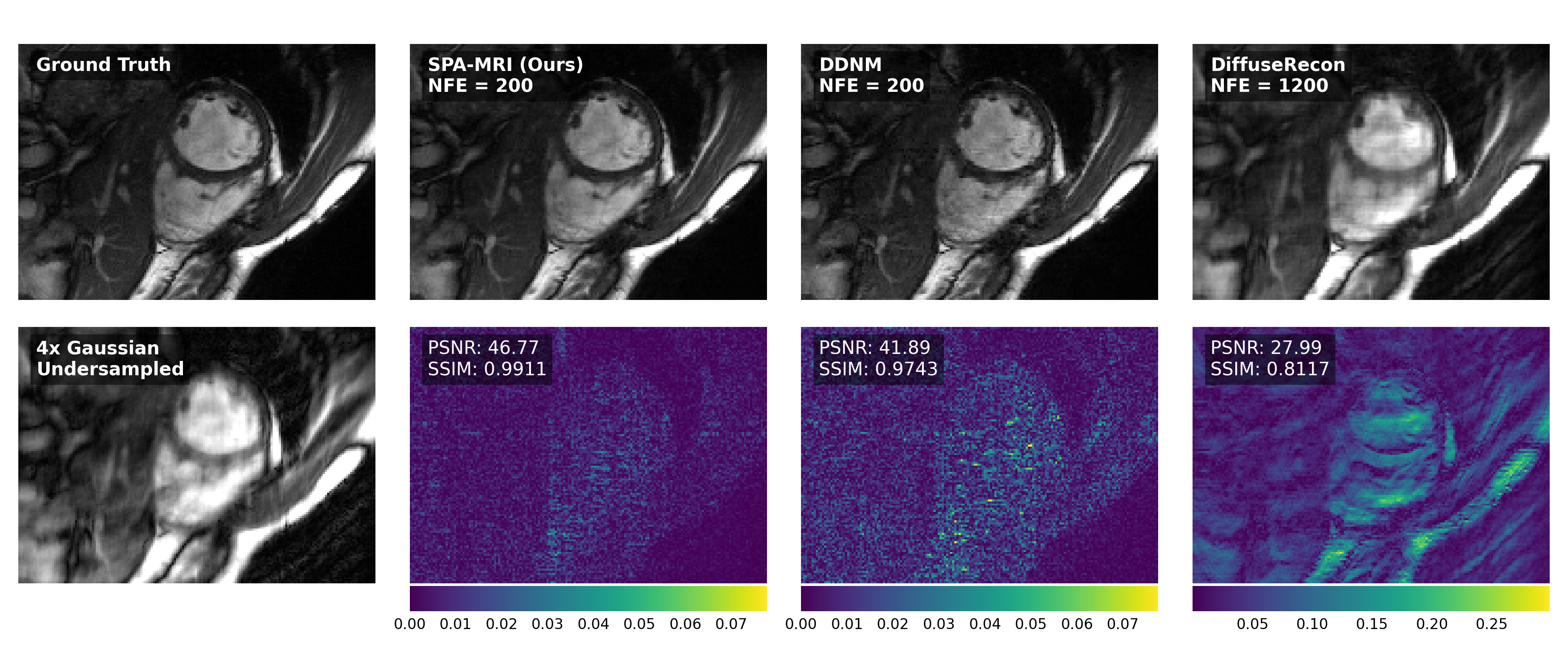}
    \caption{Comparison of MRI reconstruction methods on 4$\times$ Gaussian undersampled Cine-SAX. Here, it shows the ground truth, zero-filled image, and reconstructions of our proposed methods, DDNM\cite{wang2022zero} and DiffuseRecon\cite{Peng} along with their corresponding error maps. We also include the total Number of Function Evaluations (NFEs). For DiffuseRecon, the reconstruction is done coil-wise. The proposed approach outperforms baseline methods significantly with fewer function evaluations.}
    \label{fig:cine-recon}
\end{figure}

\vspace{+3pt}
\section{Results}
We compared our results with DiffuseRecon\cite{Peng} and DDNM\cite{wang2022zero} for 4, 12, and 24 acceleration factors with Gaussian, radial and uniform under-sampling patterns. Following challenge guidelines, we evaluated central slices, initial frames for Cine, Tagging, and Aorta, and all weightings for Mapping.   

\begin{figure}[!t]
    \centering
    \includegraphics[width=0.9\linewidth]{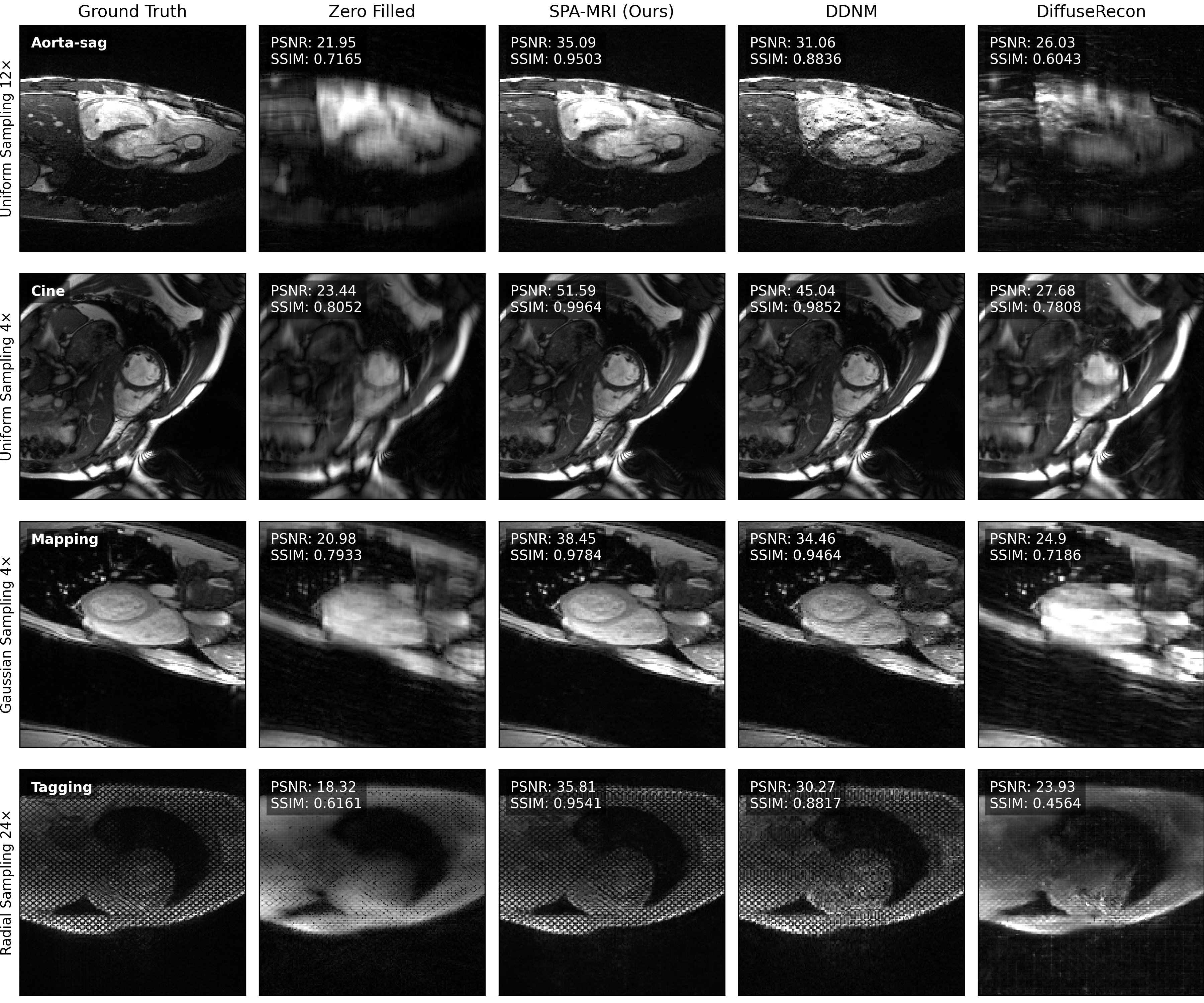}
    \caption{Comparison of MRI reconstruction across all contrasts at varying acceleration factors and undersampling patterns. Our method consistently outperforms the competing methods across Aorta, Cine, Mapping (T1 visualised), and Tagging modalities.}
    \label{fig:mm-recon}
\end{figure}

Quantitative results for Gaussian under-sampling are shown in \cref{Gaussian result}, and qualitative results are visualised in \cref{fig:cine-recon} and \cref{fig:mm-recon}. Full results on radial and uniform sampling at different acceleration factors are included in the supplementary material.  

For calculating metrics, we first normalise all the reconstructed samples to image space and scale them to $[0,1]$, then use the normalised images to compute the PSNR and SSIMs. These metrics were calculated using the following formulas:
\begin{equation}
\text{PSNR} = 20 \cdot \log_{10}(\text{MAX}_I) - 10 \cdot \log_{10}(\text{MSE}),
\end{equation}
where $\text{MAX}_I$ is the maximum possible pixel intensity (1.0 in our normalized images) and MSE is the mean squared error between the ground truth and reconstructed image.
\begin{equation}
\text{SSIM}(x,y) = \frac{(2\mu_x\mu_y + c_1)(2\sigma_{xy} + c_2)}{(\mu_x^2 + \mu_y^2 + c_1)(\sigma_x^2 + \sigma_y^2 + c_2)},
\end{equation}
where $\mu_x$ and $\mu_y$ are the average pixel intensities of images $x$ and $y$, $\sigma_x^2$ and $\sigma_y^2$ are the variances of $x$ and $y$, $\sigma_{xy}$ is the covariance of $x$ and $y$, and $c_1$ and $c_2$ are constants to stabilise the division with weak denominators.

\begin{table}[!t]
    \scriptsize
    \centering
    \renewcommand{\arraystretch}{1}
    \setlength{\tabcolsep}{0.5pt}
    \begin{tabular}{ll|c|c|c|c|c|c|c|c|c|c}
        \Xhline{1.5pt}
        \multicolumn{2}{c|}{\multirow{2}{*}{\diagbox{Acc}{Contrast}}} & \multicolumn{2}{c|}{Cine} & \multicolumn{2}{c|}{Aorta}   & \multicolumn{2}{c|}{Tagging}  & \multicolumn{2}{c|}{Mapping}  & \multicolumn{2}{c}{Total} \\
        &                                                             & PSNR $\uparrow$ & SSIM $\uparrow$  & PSNR $\uparrow$  & SSIM $\uparrow$                 & PSNR $\uparrow$  & SSIM $\uparrow$                & PSNR $\uparrow$  & SSIM $\uparrow$                & PSNR $\uparrow$  & SSIM  $\uparrow$ \\
        \hline
        \multirow{4}{*}{4$\times$} &  Zero-Filled                           & 22.64 & 0.7172            & 21.42 & 0.6324               & 21.27 & 0.6697                & 23.16 & 0.7214                & 22.12 & 0.6852 \\
                            &  DiffuseRecon                        & 31.28 & 0.7550            & 31.80 & 0.7633               &  30.85 & 0.7980               & 28.68 & 0.7605                         & 30.65 & 0.7692 \\
                            &  DDNM                                   & 40.33 & 0.9695            & 41.85 & 0.9705               &  36.66 & 0.9547               & 36.42 & 0.9231                & 38.82 & 0.9545 \\
                            &  SPA-MRI                                   & \textbf{44.73} & \textbf{0.9885}            & \textbf{46.35} & \textbf{0.9893}               & \textbf{42.45} & \textbf{0.9864}                & \textbf{41.73} & \textbf{0.9818}                & \textbf{43.82} & \textbf{0.9865} \\
        \hline
        \multirow{4}{*}{12$\times$}&  Zero-Filled                            & 21.29 & 0.6556            & 20.15 & 0.5735               & 19.13 & 0.6011                & 21.81 & 0.674                 & 20.60 & 0.6261 \\
                            &  DiffuseRecon                           & 27.86 & 0.6084            & 28.07 & 0.6115               & 26.38 & 0.6259                & 25.93 & 0.6441                         & 27.06 & 0.6225 \\
                            &  DDNM                                   & 33.98 & 0.9046            & 35.45 & 0.9074               & 31.76 & 0.8887                & 30.90 & 0.8256                & 33.02 & 0.8816 \\
                            &  SPA-MRI                                   & \textbf{38.16} & \textbf{0.9647}            & \textbf{39.35} & \textbf{0.9645}               & \textbf{36.40} & \textbf{0.9592}                 & \textbf{34.70} & \textbf{0.9434}                 & \textbf{37.15} & \textbf{0.9580} \\

        \hline
        \multirow{4}{*}{24$\times$}&  Zero-Filled                            & 21.06 & 0.6396            & 20.05 & 0.5635               & 18.94 & 0.5891              & 21.61 & 0.6669                & 20.42 & 0.6148 \\
                            &  DiffuseRecon                           & 26.74 & 0.5496            & 26.59 & 0.5451               & 25.21 & 0.5757                         & 25.09 & 0.6103                        & 25.89 & 0.5702 \\
                            &  DDNM                                   & 31.14 & 0.8649            & 32.66 & 0.843                & 29.87 & 0.8348                & 29.89 & 0.7864                & 30.89 & 0.8323 \\
                            &  SPA-MRI                                   & \textbf{35.02} & \textbf{0.9434}            & \textbf{37.46} & \textbf{0.9517}               & \textbf{33.69} & \textbf{0.9375}                & \textbf{33.79} & \textbf{0.9322}                & \textbf{34.99} & \textbf{0.9412} \\
        \Xhline{1.5pt}
        \end{tabular}
        \vspace{+5pt}
        \caption{Numerical results of zero-filled, DiffuseRecon\cite{Peng}, DDNM\cite{wang2022zero} and our proposed methods for Gaussian undersample mask with different acceleration factors. The best results are in \textbf{bold}. Metrics were calculated on the full reconstructed images.}
        \label{Gaussian result} 
\end{table}

The results demonstrate that the proposed SPA-MRI method consistently outperforms baseline methods across different MRI reconstruction scenarios. From \cref{Gaussian result}, we can observe that SPA-MRI achieves the highest PSNR and SSIM scores for all acceleration factors across all contrast types. The performance gap is particularly notable at higher acceleration factors, indicating the robustness of the frequency decomposed loss to higher accelerations.

SPA-MRI achieves faster reconstruction times compared to DDNM while maintaining superior image quality. The slight increase in computation time relative to DiffuseRecon is offset by the significant improvement in reconstruction accuracy. Future work will focus on further optimizing the algorithm for real-time applications.

\cref{fig:cine-recon} further visually confirms the results, where the proposed SPA-MRI produces fewer artefacts and better preserves fine details compared to DDNM and DiffuseRecon whilst having fewer inference steps. We use the same underlying UNet model \cite{ronneberger2015unet} for sampling; thus, we compare the required sampling steps (NFEs) as well in \cref{fig:cine-recon}, the proposed method effectively reconstructs with higher PSNR and SSIM while using the least steps.

\cref{fig:mm-recon} consistently extends these observations across multiple MRI contrasts with varying acceleration factors and undersampling patterns. Note that the model has been trained jointly on all modalities without separate training on each of them, and the consistency enforcement mechanism has further enabled the model to deal with arbitrary undersampling patterns and acceleration factors. This generalisability is a key strength of the proposed method.  

\vspace{+5pt}
\noindent\textbf{Reconstruction Time Analysis.}
\begin{table}[!t]
    \centering
    \begin{tabular}{l|c|c}
        \hline
        Method        & NFE & Time/Slice (s) \\
        \hline
        DiffuseRecon  & 1200 & 285.3  \\
        DDNM          & 200  & 44.15  \\
        SPA-MRI       & 200* & 40.04  \\
        \hline
    \end{tabular}
    \vspace{+5pt}
    \caption{Comparison of total number of function Evaluations (NFEs) and average reconstruction time per slice (in seconds) using the same underlying diffusion model. Note that for SPA-MRI, the total NFEs include the inversion and the sampling steps.}
    \label{tab:recon_time}
\end{table}
To evaluate the computational efficiency of SPA-MRI, we conducted timing experiments on a single NVIDIA A5000 GPU. Table \ref{tab:recon_time} presents the average reconstruction time per slice for SPA-MRI compared to baseline methods. The SPA-MRI method achieves high PSNR and SSIM scores while requiring significantly fewer function evaluations than DiffuseRecon and DDNM. In our results, we consider the extra sampling steps necessary for the diffusion-aware inversion process. Despite these additional steps, our enhanced seed initialisation enables us to reduce the overall sampling steps, leading to faster inference.

\vspace{+5pt}
\noindent\textbf{Limitations and Future Work.}
Our method achieves significantly better results than baseline methods; however, certain limitations may hinder its broader applicability. First, our approach ignores the noise components in the acquisition process (see \cref{y_x_relation in MRI}), which is different from real-world scenarios. Including the noise models into the sampling process is a promising scope for future work and can potentially result in more accurate reconstructions.   
Besides, training a diffusion model is more computationally intensive than other traditional CNN-based methods. Our diffusion UNet \cite{ronneberger2015unet} comprises 150 million parameters, whereas contemporary CNN-based methods typically range between 1-10 million parameters. This could limit its practicality due to the requirement for large GPUs during inference. While our inference times are competitive, additional optimization would be necessary for real-world applications. Model distillation and parameter-efficient adaption can further increase the applicability of the proposed methods.

\section{Conclusion}

In this paper, we proposed SPA-MRI, a diffusion-based multi-coil MRI reconstruction model that generalises across contrasts, acceleration factors, and sampling patterns. We train an unconditional diffusion model and reconstruct MR images with adaptive consistency enforcement. By combining frequency decomposed posterior loss with adaptive weighing for enforcing consistency, SPA-MRI achieves superior performance across modalities, undersampling patterns and acceleration factors, facilitating broader deployment of accelerated MRI without any downstream fine-tuning. 

\section*{Acknowledgement}
This work was supported by the Engineering and Physical Sciences Research Council (EPSRC) UK grant TrustMRI [EP/X039277/1]. C. Wang was supported by the Shanghai Municipal Science and Technology Major Project (No. 2023SHZD2X02A05), the National Natural Science Foundation of China (No. 62331021) and the Shanghai Sailing Program (No. 20YF1402400).

\bibliographystyle{splncs04}
\bibliography{bibliography}
\end{document}

%% file: methodology.tex
\vspace{-10pt}
\subsubsection{Enhancement of the Forward Noising Process} 
\begin{figure}[!t]
    \centering
    \begin{subfigure}{.5\textwidth}
      \centering
      \includegraphics[width=\linewidth]{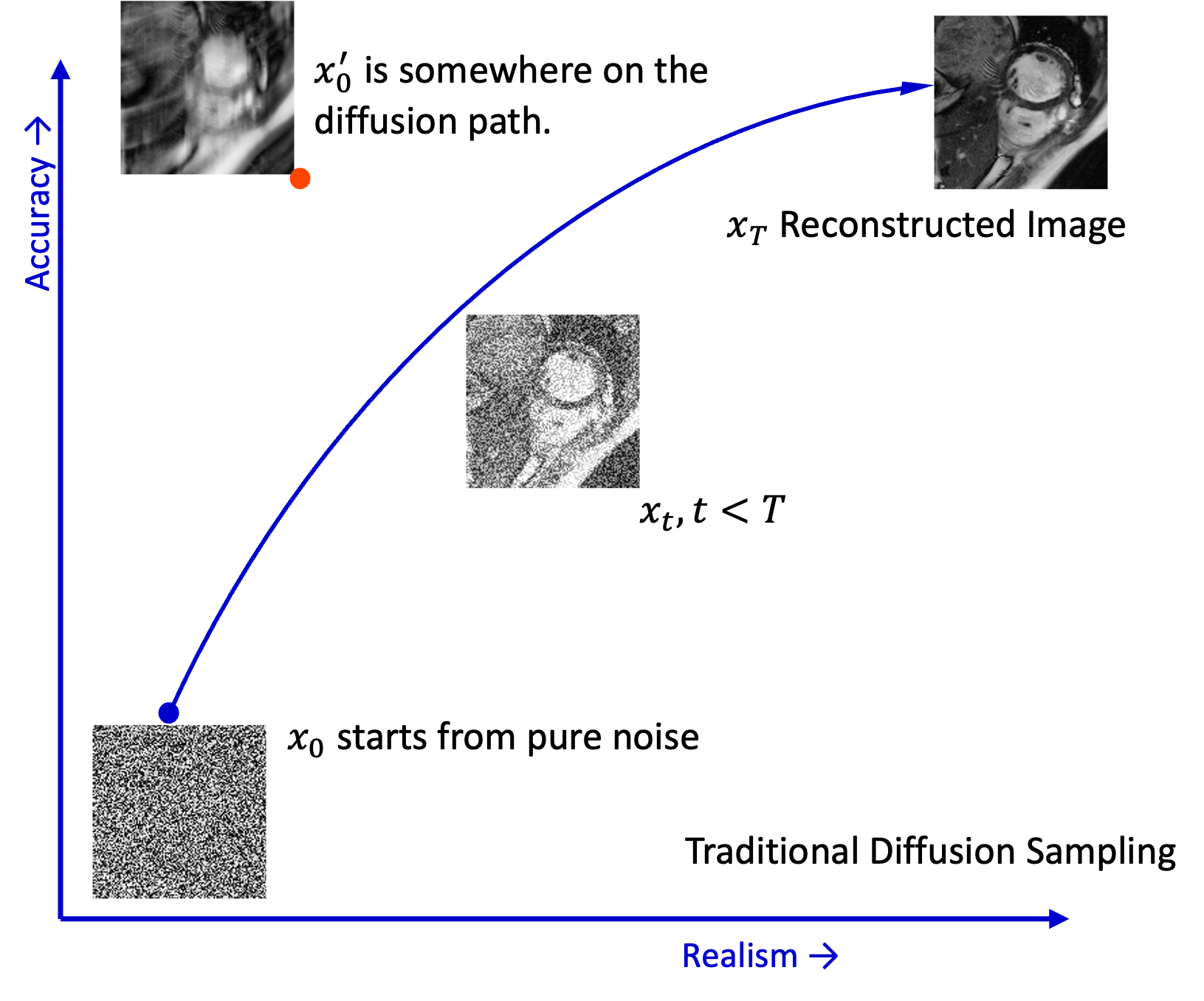}
      \caption{Standard Diffusion Sampling}
      \label{fig:sub1}
    \end{subfigure}%
    \begin{subfigure}{.5\textwidth}
      \centering
      \includegraphics[width=\linewidth]{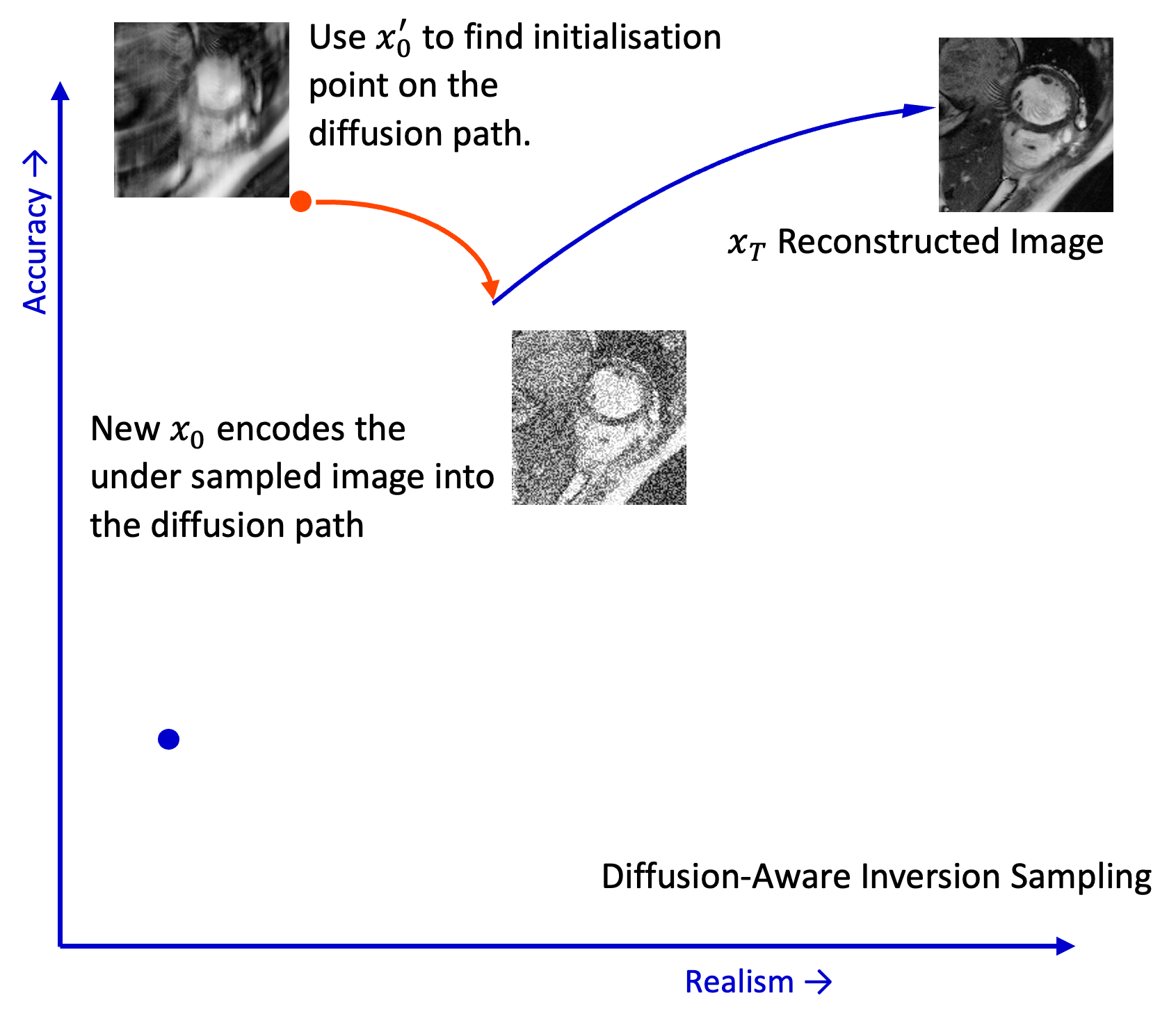}
      \caption{Diffusion Aware Inversion}
      \label{fig:sub2}
    \end{subfigure}
    \caption{Conceptual visualisation of the typical diffusion sampling process (a) is compared with the enhanced diffusion-aware inversion proposed by Liu et al. (b) \cite{liu2023accelerating}. The inversion process comprises two stages: first, it adjusts the undersampled image $x_0'$ to the diffusion process trajectory, and then it follows the standard diffusion process, requiring significantly fewer sampling steps.}
    \label{fig:DAI}
\end{figure}

To overcome the limitation of initialisation in existing approaches, inspired by Distortion Adaptive Inversion \cite{liu2023accelerating}, we propose to follow an improved diffusion-aware forward noising process (Eq. \ref{dai-forward}) to generate the initial noise seed:
\begin{equation}
    \label{dai-forward}
            \mathbf{x}_{t+1} = \sqrt{\alpha_{t+1}} \left( \mathbf{x}_0(\mathbf{x}_t)\epsilon_\theta(\mathbf{x}_t, t) \right) 
                     + \sqrt{1-\alpha_{t+1} - \beta_{t+1}}\epsilon_\theta(\mathbf{x}_t, t)
                     + \sqrt{\beta_{t+1}}\mathbf{z}.
\end{equation}
Here the initialisation is modified to include controllable perturbations $\mathbf{z} \sim \mathcal{N}(0, \mathbf{I})$ for preserving essential information. The layout and structure are shown in \cref{fig:DAI}.   

With this optimised initialisation, this approach can significantly reduce the effective denoising steps needed (i.e., function evaluations) while performing the reconstruction from the under-sampled data. It is also shown that the method can better preserve reconstruction realism and distribution fidelity.

\vspace{-10pt}
\subsubsection{Integration of Data-Consistency for Parallel Imaging}

Incorporating data consistency for parallel imaging, MR is non-trivial. While methods like DPS \cite{chungdps}, DDNM \cite{wang2022zero} and DiffuseRecon \cite{Peng} work well for single-coil data, they fail to leverage parallel imaging information.   


In our method, we extend DDNM's back projection method for parallel imaging. We decompose $\mathbf{x}_0(\mathbf{x}_t)$ into coil-wise estimates, perform consistency per coil, then project errors back to combined space: $\mathcal{A}^T(\mathcal{A}(\mathbf{x}_0(\mathbf{x}_t))^c - \mathbf{y}^c)$, where $c \in [0, C]$ represents coil-wise measurements.  
Our approach incorporates multi-coil information without the costly optimisation steps like CG, as seen in \cite{chung2023dds} and \cite{aggarwal2018modl}. While CG methods may yield higher fidelity samples, our method achieves sufficiently good results more efficiently.  

We have observed that the magnitude of the back projected errors change over time and benefits from scaling their influence, resulting in higher quality reconstructions. The scaling balances between greater influence at larger error scales for faster convergence and finer control when the errors are small to promote refined adjustments to $\mathbf{x}_0(\mathbf{x}_t)$. Therefore, we introduce an adaptive weighting schema for the consistency step:
\begin{equation}
    \mathbf{x}_0(\mathbf{x}_t) = \mathbf{x}_0(\mathbf{x}_t) -  \omega \mathcal{A}^T(\mathcal{A}(\mathbf{x}_0(\mathbf{x}_t)) - \mathbf{y}),
    \label{adaptive-bp}
\end{equation}
where $\omega = \xi * (1+\tanh(\delta_{t+1} - \delta_t)/2)$, $\xi$ is the base scale, and $\delta_t = ||\Delta(\mathbf{x}_0(\mathbf{x}_t), \mathbf{y)||_2}$ (see \cref{sec:freq}) is the error magnitudes at time $t$. Using the hyperbolic tangent function, we scale the difference to $[-1, 1]$. This effectively adapts $\xi$ to increase or decrease based on the error changes. 
\vspace{-10pt}
\subsubsection{Frequency Decomposed Posterior Loss}
\label{sec:freq}

Motivated by the distinct characteristics of low- and high-frequency information in MRI data, we propose a novel decomposition of the back-projection consistency measurement $\Delta$:
\begin{equation}
    \label{eq:bp_err}
    \Delta(\mathbf{x}, \mathbf{y}) = \mathcal{A}(\mathbf{x}) - \mathbf{y}.
\end{equation}
This approach, largely unexplored in consistency space, leverages the guaranteed accuracy of low-frequency Auto-Calibration Signal (ACS) data while focusing on reconstructing challenging high-frequency details, improving overall reconstruction quality and accuracy:
\begin{equation}
\Delta_{\text{d}}(\mathbf{x}_0(\mathbf{x}_t), \mathbf{y}) = \lambda_1 \Delta_{\text{low}}(\mathbf{x}_0(\mathbf{x}_t), \mathbf{y}) + \lambda_2 \Delta_{\text{high}}(\mathbf{x}_0(\mathbf{x}_t), \mathbf{y}).
\label{freq_decomp}
\end{equation}
Here $\lambda_1$ and $\lambda_2$ are weighting coefficients for low-frequency and high-frequency components, respectively. Specifically, $\lambda_2 > \lambda_1$, which prioritises high-frequency reconstruction. This approach is able to enhance fine detail recovery while maintaining low-frequency information integrity. 
Our proposed method is detailed in Algorithm \ref{alg:mri_reconstruction}.
 
\begin{algorithm}[!t]
\scriptsize

\caption{SPA-MRI Sampling}
\label{alg:mri_reconstruction}
\begin{algorithmic}

\For{$s = 1$ to $S$}
    \Comment{Inversion Step}
    \State $\epsilon \gets \epsilon_\theta(\mathbf{x}_s, s)$
    \State $\mathbf{x}_0(\mathbf{x}_s) \gets \frac{\mathbf{x}_s - \sqrt{1 - \alpha_s}\epsilon}{\sqrt{\alpha_s}}$
    \State $z \sim \mathcal{N}(0, \mathcal{I})$
    \State $\mathbf{x}_{s+1} = \sqrt{\alpha_{s+1}}\mathbf{x}_0(\mathbf{x}_s) + \sqrt{1-\alpha_{s+1}-\lambda\beta_{s+1}}\epsilon + \sqrt{\lambda \beta_{s+1}}z$
\EndFor
\State $\delta_T = 0$
\For{$t = T$ to $0$} \Comment{Reverse Diffusion Step}
    \State $\epsilon \gets \epsilon_\theta(x_t, t)$
    \State $\mathbf{x}_0(\mathbf{x}_t) \gets \frac{\mathbf{x}_t - \sqrt{1 - \alpha_t}\epsilon_2}{\sqrt{\alpha_t}}$
    \State $\delta_t \gets || \Delta_{d}(\mathbf{x}_0(\mathbf{x}_t), \mathbf{y}) ||_2$
    \State $\mathbf{x}_0(\mathbf{x}_t) \gets \mathbf{x}_0(\mathbf{x}_t) - \xi (1 + \tanh (\delta_{t+1} - \delta_t)/2) \mathcal{A}^T(\Delta_{d}(\mathbf{x}_0(\mathbf{x}_t), \mathbf{y}))$
    \State $x_{t-1} \gets \sqrt{\alpha_{t-1}}x_0(x_t) + \sqrt{1-\alpha_{t-1}-\sigma^2}\epsilon_2 + \sigma z$ \Comment{$\sigma = \frac{1-\Bar{\alpha}_{t-1}}{1-\Bar{\alpha}_t}\beta_t$}
\EndFor

\State \Return $\hat{\mathbf{x}}_0 \leftarrow \mathbf{x}_0$
\end{algorithmic}
\end{algorithm}